\documentclass[aps,pra,prabib,amsmath,amssymb,amsfonts,epsfig,smallcaptions,showpacs]{revtex4}
\usepackage{psfig}
\usepackage{graphicx}
\newcommand{\eqname}[1]{\label{eq:#1}}
\newcommand{\bgar}{\begin{eqnarray}}
\newcommand{\enar}[1]{\label{eq:#1}\end{eqnarray}}

\newcommand{\ket}[1]{ | #1 \rangle }
\newcommand{\bra}[1]{\langle  #1 |}
\newcommand{\braket}[2]{ \langle #1 | #2 \rangle }
\newcommand{\braopket}[3]{ \left\langle #1 \right| #2 \left| #3 \right\rangle }

\newcommand{\expect}[1]{\Big\langle #1 \Big\rangle}

\newcommand{\kk}{ {\bf k}}
\newcommand{\rr}{ {\bf r}}

\newcommand{\uu}{ {\bf u}}
\newcommand{\nn}{ {\bf n}}
\newcommand{\ee}{ {\bf e}}

\newcommand{\eq}[1]{(\ref{eq:#1})}
\newcommand{\al}[1]{^{(#1)}}
\newcommand{\Psihd}{\hat\Psi^\dagger}

\newcommand{\Psih}{\hat\Psi}

\begin{document}

\title{An exact reformulation of the Bose-Hubbard model in terms of
a stochastic Gutzwiller ansatz}

\affiliation{Laboratoire Kastler Brossel, Ecole Normale
Sup\'erieure, 24 rue Lhomond, 75231 Paris Cedex 05, France}

\author{Iacopo Carusotto}
\email{Iacopo.Carusotto@lkb.ens.fr}
\affiliation{Laboratoire Kastler Brossel, Ecole Normale
Sup\'erieure, 24 rue Lhomond, 75231 Paris Cedex 05, France}

\author{Yvan Castin}
\affiliation{Laboratoire Kastler Brossel, Ecole Normale
Sup\'erieure, 24 rue Lhomond, 75231 Paris Cedex 05, France}

\begin{abstract}
We extend our exact reformulation of the bosonic many-body problem
in terms of a stochastic Hartree ansatz to a stochastic Gutzwiller ansatz for
the Bose-Hubbard model. The use of this ansatz makes the corresponding Monte
Carlo scheme more 
efficient for strongly correlated bosonic phases like the Mott
insulator phase or the Tonks phase. We present a first numerical 
application of this stochastic
method to a system of impenetrable bosons on a 1D lattice showing the
transition from the discrete Tonks gas to the Mott phase as the chemical potential
is increased.
\end{abstract}

%\vspace{1cm}

% (Boson systems)
%(Phase coherent atomic ensembles; quantum condensation phenomena)
%(Quantum Monte Carlo methods)

\date{\today}

\maketitle

\section{Introduction}

The Bose-Hubbard model of lattice bosons with on-site interactions~\cite{BoseHubbard} is
a widely used theoretical model for the description of a number of different
physical systems, ranging from $^4$He adsorbed in porous
media~\cite{HePorous} to granular
superconductors~\cite{GranularSupercond,GranularSupercond2} and Josephson junction
arrays~\cite{Josephson,Josephson2,Josephson3,Josephson4,Josephson5,Josephson6}. 

In the last years, the Bose-Hubbard model has attracted a renewed interest since its
experimental realisation with ultracold atoms loaded in the periodic
potential of an optical lattice \cite{OptLattices,OptLattices2}. Following the
suggestion of \cite{Zoller}, this system has led to the observation of the superfluid-Mott insulator
quantum phase transition \cite{OptLattices2} and of quantum revivals of the matter field
\cite{PhaseRev}.  In the present paper we propose a new tool for the study of
the Bose-Hubbard model in the form of a Quantum Monte Carlo scheme based
on the stochastic evolution of a Gutzwiller ansatz. This new scheme
generalizes our previously proposed schemes based on a stochastic Hartree
ansatz~\cite{GPstoch,GPstochT,Pedag}.

Intuitively, it is clear that the Hartree ansatz is not the optimal
one for the 
study of a Mott insulator state:
the expansion of a Mott-insulator state on Hartree states involves an
average over all the possible relative phases between adjacent sites,
so that a large number of realizations is required for an accurate
sampling of the state. Furthermore, if one rises the on-site
interaction strength to infinity so to realize the so-called
impenetrable boson model, the rate of increase of the statistical
error with time, being proportional to the strength of the atom-atom
interactions, will diverge. 

On the other hand, approximate analytical mean-field calculations
based on a Gutzwiller-type wavefunction have given predictions
which are in reasonable agreement with the exact numerical calculations \cite{KrauthGutz}.
This suggests that a suitable stochastic generalization of the
Gutzwiller mean-field equations could lead to an efficient quantum
Monte Carlo scheme. The main difference with respect to the reformulation in
terms of a stochastic Hartree ansatz consists in the exchanged roles
of the kinetic and the interaction energies: in the Gutzwiller case,
interactions can be fully taken into account by the deterministic evolution, while the
noise is necessary to keep track of the correlations between different
sites which are induced by the kinetic energy term in the Hamiltonian.
While the stochastic Hartree reformulation was well suited for the study of weakly
interacting Bose gases~\cite{StatN0}, the present Gutzwiller scheme is expected to
be more suited to the study of systems of strongly interacting bosons,
such as the Mott phase or the gas of impenetrable bosons.

In sec.\ref{sec:Gutz} we review the Bose-Hubbard Hamiltonian and we
introduce the Gutzwiller ansatz. In sec.\ref{sec:TimeEvol} 
we show how it is possible to find a suitable stochastic evolution of the Gutzwiller
wave function such that the stochastic evolution of the ansatz is
equal, in the average, to the exact evolution given by the Bose-Hubbard
Hamiltonian \eq{Hamilt}. 
In sec.\ref{sec:StatErr} we discuss the
evolution of the statistical error of this stochastic reformulation of
the many-body problem and, in particular, we show that the error
remains finite at any time of the evolution. This guarantees that the
stochastic Gutzwiller reformulation can be actually used for numerical Monte Carlo calculations.
In sec.\ref{sec:ImagTime}, we extend the reformulation to the case of an
imaginary-time evolution so as to obtain the thermal equilibrium state
of the Bose-Hubbard system at a given temperature and chemical
potential. A first application to the case of impenetrable bosons on a
1D lattice is presented, which shows the transition from a
discrete Tonks gas to the Mott phase as the chemical potential is
increased: the results are in good agreement with analytical calculations
performed by means of fermionization techniques. For the sake of
completeness, a brief discussion of these techniques is given in the appendix. 
In sec.\ref{sec:Spin} we generalize the results to the case of several
spin components. Conclusions are drawn in sec.\ref{sec:Conclu}.

\section{The model Hamiltonian and the Gutzwiller ansatz}
\label{sec:Gutz}

In its simplest form, the Bose-Hubbard Hamiltonian can be written as:
\begin{equation}
\label{eq:Hamilt}
{\mathcal H}={\mathcal H}_U+{\mathcal H}_J=\frac{U}{2}\,\sum_\rr
\Psihd(\rr)\Psihd(\rr)\Psih(\rr)\Psih(\rr)+\frac{J}{2}\,\sum_{\rr,\rr'}\,\delta_{\rr,\rr'}\al{1}\Psihd(\rr)\Psih(\rr'),
\end{equation}
in which the spatial coordinate $\rr$ runs on a $D $-dimensional
cubic lattice of ${\mathcal N}=\prod_{i=1}^D {\mathcal N}_i$ points with periodic
boundary conditions. With $l_i$ and $L_i={\mathcal N}_i l_i$, we shall denote
respectively the lattice spacing and the total lattice size along the $i$-th 
dimension.
The on-site interaction is taken into account by the first term ${\mathcal
H}_U$, which is proportional to the charging energy $U$. 
The hopping from one site to a neighboring one is described by the other term
${\mathcal H}_J$. $J$ quantifies the amplitude of the hopping processes.
The function $\delta\al{1}_{\rr,\rr'}$ is 
equal to one if $\rr$ and $\rr'$ are neighboring sites, otherwise it vanishes.
The annihilation operators $\Psih(\rr)$ satisfy the bosonic commutation rules
$[\Psih(\rr),\Psihd(\rr')]=\delta_{\rr,\rr'}$.

The generic Gutzwiller ansatz can be written in terms of the complex
wavefunctions $\phi(\rr,n)$ as:
\begin{equation}
\ket{G:\phi}=\prod_\rr{\sum_{n=0}^{\infty}
  \phi(\rr,n) \frac{\Psihd(\rr)^n}{\sqrt{n!}}}\,\ket{0}.
\label{eq:Gutzwiller}
\end{equation}
This kind of states has been extensively used within mean-field theory
to obtain approximate analytical results for strongly interacting bosons on a
lattice: since the wave function has a factorized form, no correlations between
different sites are included and the effect of ${\mathcal H}_J$ can
not be treated exactly. This is complementary to what happens in the
case of a Hartree ansatz where the mean field theory can deal exactly
with the hopping Hamiltonian, but not with the interaction
one.

The generality of the Gutzwiller ansatz is however large enough to include
Glauber coherent states as well as Fock states. A coherent state 
of macroscopic wavefunction $\gamma(\rr)$:
\begin{equation}
\ket{\textrm{coh}:\gamma}=e^{-\|\gamma\|^2/2}\,\sum_{n=0}^\infty\,\frac{1}{\sqrt{n!}}
\Big(\sum_\rr\gamma(\rr)\Psihd(\rr)\Big)^n\,\ket{0},
\eqname{Coherent}
\end{equation}
where $\|\gamma\|^2=\sum_{\rr} |\gamma(\rr)|^2$, corresponds to a Gutzwiller
state with:
\begin{equation}
\phi(\rr,n)=e^{-|\gamma(\rr)|^2/2}\,\frac{\gamma(\rr)^n}{\sqrt{n!}}.
\eqname{GutzCoh}
\end{equation}
On the other hand, a Fock state in which each site has a well defined
occupation $n_0(\rr)$ corresponds to a Gutzwiller wavefunction of the form:
\begin{equation}
\phi(\rr,n)=\delta_{n,n_0(\rr)}.
\end{equation}
The Gutzwiller ansatz is therefore much more flexible than the coherent state
ansatz often used in quantum optics.
Its normalization is given by:
\begin{equation}
\braket{G:\phi_g}{G:\phi_d}=\prod_\rr\sum_{n=0}^\infty \phi_g^*(\rr,n)\phi_d(\rr,n).
\label{eq:GutzNormal}
\end{equation}

\section{Real time evolution}
\label{sec:TimeEvol}
As the set of all Gutzwiller states forms an overcomplete family,
one can always assume that the initial state vector  $\ket{\psi(0)}$ is a
coherent superposition of Gutzwiller states. 
As discussed in~\cite{Pedag} for the case of a Hartree ansatz, one can
reabsorb the phases of 
the complex coefficients of this superposition by a redefinition of the
Gutzwiller wavefunctions, so that the initial 
state vector can be written as the average of the Gutzwiller
ansatz over some probability distribution for $\phi$:
\begin{equation}
\ket{\psi(0)}=\Big\langle\;\ket{G:\phi}\;\Big\rangle
\eqname{Average0}
\end{equation}
where we represent this average by angular brackets $\big\langle\ldots\big\rangle$.
In the present section, we shall show that a suitable stochastic evolution
can be found for the Gutzwiller wavefunctions $\phi(\rr,n)$, such that a
relation like \eq{Average0} holds at any time $t>0$, in the
sense that the exact state vector at time $t$ is given by the
average over the ensemble of Gutzwiller states resulting from all possible stochastic
evolutions of the initial ensemble during the time $t$:
\begin{equation}
\ket{\psi(t)}=\Big\langle\;\ket{G:\phi(t)}\;\Big\rangle.
\eqname{Average}
\end{equation}

In the hopping Hamiltonian ${\mathcal H}_J$, we first explicitely isolate the
mean-field contribution by rewriting:
\begin{equation}
{\mathcal H}_J=\frac{J}{2}\,\sum_{\rr,\rr'}
\,\delta_{\rr,\rr'}\al{1}\Psihd(\rr)\Psih(\rr')=\frac{J}{2}\,\sum_{\rr,\rr'}
\,\delta_{\rr,\rr'}\al{1}\left[(\Psihd(\rr)-{\bar \psi}^*(\rr))
(\Psih(\rr')-{\bar \psi}(\rr'))+
\Psih(\rr)\,{\bar \psi}^*(\rr')+\Psihd(\rr)\,{\bar \psi}(\rr')-
{\bar \psi}^*(\rr)\,{\bar \psi}(\rr')\right]
\eqname{H_J}
\end{equation}
in terms of the ${\mathbf C}$-number mean-field ${\bar \psi}(\rr)$.
defined  as:
\begin{equation}
{\bar \psi}(\rr)=\frac{\braopket{G:\phi}{\Psih(\rr)}{G:\phi}}
{\braket{G:\phi}{G:\phi}}=\frac{\braopket{\phi(\rr)}{\Psih}{\phi(\rr)}}{\braket{\phi(\rr)}{\phi(\rr)}}
\end{equation}
where the ket $\ket{\phi(\rr)}$ describes the state of the site $\rr$:
\begin{equation}
\ket{\phi(\rr)}=\sum_{n=0}^\infty \phi(\rr,n)\,\ket{n}.
\end{equation}
This corresponds to the general strategy, already implemented for the
Hartree-Fock ansatz for bosons \cite{Pedag} and for fermions \cite{Chomaz}, to have
a deterministic evolution identical to the mean-field theory so that
the stochastic part includes only deviations from mean-field.
Mathematically, as we shall see, this makes the noise term strictly
orthogonal to the state $|G:\phi\rangle$, which is expected to
minimize the growth rate 
of the statistical error.
Note that only the first term of the square bracket \eq{H_J} can induce correlations
between different sites, and it is this term which will be taken into account with
the noise term of the evolution.

The action of the total Hamiltonian \eq{Hamilt} on a Gutzwiller state
of the form \eq{Gutzwiller} can then be written as:
\begin{multline}
\frac{1}{i\hbar}{\mathcal H}\,\ket{G:\phi}\,dt=\sum_\rr
\Big[\sum_{n,n'=0}^\infty
  \chi(\rr;n,n')\,\phi(\rr,n')\,\frac{\Psihd(\rr)^n}{\sqrt{n!}}  
\Big]\, \Big[ \prod_{\uu\neq\rr}
  \sum_{n=0}^\infty\phi(\uu,n)\frac{\Psihd(\uu)^n}{\sqrt{n!}}
\Big]\,\ket{0}+\\+
\frac{J\,dt}{2i\hbar}\sum_{\rr,\rr'}\delta\al{1}_{\rr,\rr'}\Big[\sum_{n,n'=0}^\infty
\Big(\sqrt{n}\,\phi(\rr,n-1)-{\bar \psi}^*(\rr)\,\phi(\rr,n)\Big)
\Big(\sqrt{n'+1}\,\phi(\rr',n'+1)-{\bar \psi}(\rr')\,\phi(\rr',n')\Big)
\,\frac{\Psihd(\rr)^n\,\Psihd(\rr')^{n'}}{\sqrt{n!\,n'!}} \Big]\cdot \\
\cdot\Big[\prod_{\uu\neq\rr,\rr'} 
  \sum_{n=0}^\infty\phi(\uu,n)\frac{\Psihd(\uu)^n}{\sqrt{n!}} \Big]\,\ket{0},
\eqname{evol_H}
\end{multline}
where the matrices $\chi(\rr;n,n')$ have vanishing entries for
$|n-n'|>1$,
\begin{equation}
  \label{eq:chi0}
  \chi(\rr;n,n)=\frac{dt}{i\hbar}\Big[\frac{U\,n(n-1)}{2}-\frac{J}{2{\mathcal N}}\sum_{\uu,\uu'}\delta\al{1}_{\uu,\uu'}{\bar \psi}^*(\uu)\,{\bar \psi}(\uu')\Big],
\end{equation}
\begin{equation}
  \label{eq:chi1}
  \chi(\rr;n,n+1)=\frac{J\,dt}{2i\hbar}\,\sqrt{n+1}\,\sum_{\rr'} \delta\al{1}_{\rr,\rr'}
{\bar \psi}^*(\rr')
\end{equation}
and
\begin{equation}
  \label{eq:chi2}
  \chi(\rr;n,n-1)=\frac{J\,dt}{2i\hbar}\,\sqrt{n}\,\sum_{\rr'} \delta\al{1}_{\rr,\rr'}
{\bar \psi}(\rr'),
\end{equation}
%\begin{multline}
%  \label{eq:chi}
%  \chi(\rr;n,n')=\\=\frac{dt}{i\hbar}\Big[\frac{U\,n(n-1)}{2}\delta_{n,n'}+
%\frac{J}{2}\sum_{\rr'} \delta\al{1}_{\rr,\rr'}\Big({\bar
%  \psi}(\rr')\,\sqrt{n}\,\delta_{n-1,n'}+
%{\bar
%  \psi}^*(\rr')\,\sqrt{n+1}\,\delta_{n+1,n'}\Big)
%-\frac{J}{2{\mathcal N}}\sum_{\uu,\uu'}\delta\al{1}_{\uu,\uu'}{\bar \psi}^*(\uu)\,{\bar
%  \psi}(\uu')\,\delta_{n,n'}\Big]
%\end{multline}
so to include all the on-site terms coming from both the interaction and 
the mean-field part of the hopping Hamiltonian \eq{H_J}.
The factor $1/{\mathcal N}$ appearing in the last term in \eq{chi0} comes from the
normalization \eq{GutzNormal} of the Gutzwiller state.

We now consider a generic stochastic evolution for $\phi(\rr,n)$:
\begin{equation}
d\phi(\rr,n)=d\phi_d(\rr,n)+d\phi_s(\rr,n)
\end{equation}
where $d\phi_d(\rr,n)$ is a deterministic drift term proportional to $dt$ and
$d\phi_s(\rr,n)$ is a stochastic noise in the Ito sense. In particular, the
mean value of $d\phi_s$ is zero and its variance proportional to $dt$.
The average evolution of the Gutzwiller ansatz during $dt$ can be obtained by
means of an expansion in powers of $dt$; retaining the terms up to order $dt$,
one is left with:
\begin{multline}
\overline{\ket{G:\phi+d\phi}}=\ket{G:\phi}+
\sum_\rr \Big[\sum_{n=0}^\infty
 d\phi_d(\rr,n)\frac{\Psihd(\rr)^n}{\sqrt{n!}}\Big]\,\Big[\prod_{\uu\neq\rr}
\sum_{n=0}^\infty\phi(\uu,n)\frac{\Psihd(\uu)^n}{\sqrt{n!}} 
 \Big]\, 
\ket{0} \\ 
+\frac{1}{2}\sum_{{\rr,\rr'},{\rr\neq\rr'}}\Big[\sum_{n,n'=0}^\infty
\overline{d\phi_s(\rr,n)\,d\phi_s(\rr',n')} 
\frac{\Psihd(\rr)^n\,\Psihd(\rr')^{n'}}{\sqrt{n!\,n'!}}\Big]\,
\Big[\prod_{\uu\neq\rr,\rr'}
\sum_{n=0}^\infty \phi(\uu,n)\frac{\Psihd(\uu)^n}{\sqrt{n!}}\Big]
\,\ket{0}
\eqname{evol_st}
\end{multline}
The stochastic evolution \eq{evol_st} can be made to coincide with the exact
one given by \eq{evol_H} simply by imposing that the deterministic evolution
$d\phi_d$ is equal to:
\begin{multline}
d\phi_d(\rr,n)=\sum_{n'} \chi(\rr;n,n')\phi(\rr,n')=\frac{dt}{i\hbar}\left[
\frac{U\,n(n-1)}{2}\phi(\rr,n)+ \right.\\ 
    \left.+\frac{J}{2}\sum_{\rr'}\delta_{\rr,\rr'}\al{1}
    \left({\bar
    \psi}(\rr')\,\sqrt{n}\,\phi(\rr,n-1) +{\bar
    \psi}^*(\rr')\,\sqrt{n+1}\,\phi(\rr,n+1)\right) 
    -\frac{J}{2{\mathcal N}}\Big(\sum_{\uu,\uu'}\delta_{\uu,\uu'}\al{1} {\bar
    \psi}^*(\uu){\bar \psi}(\uu')\Big)\,\phi(\rr,n)\right]
\eqname{MFd}
\end{multline}
and that the correlation function of the noise term satisfies:
\begin{multline}
\overline{d\phi_s(\rr,n)\,d\phi_s(\rr',n')}
=\frac{J\,dt}{2i\hbar}\delta_{\rr,\rr'}\al{1}\Big[\big(\sqrt{n}\,\phi(\rr,n-1)-{\bar
  \psi}^*(\rr)\,\phi(\rr,n) \big)\big(\sqrt{n'+1}\,\phi(\rr',n'+1)-{\bar
  \psi}(\rr')\,\phi(\rr',n')\big)+\\+\big(\sqrt{n+1}\,\phi(\rr,n+1)-{\bar
  \psi}(\rr)\,\phi(\rr,n) \big)\big(\sqrt{n'}\phi(\rr',n'-1)-{\bar
  \psi}^*(\rr')\,\phi(\rr',n')\big)\Big].
\eqname{NoiseCorr}
\end{multline}
The symmetrized form of \eq{NoiseCorr} with respect to $(\rr,n)$ and
$(\rr',n')$ is imposed by the general property of a noise correlation
matrix being symmetric.
Apart from an additional term giving a mere phase rotation, the deterministic
evolution \eq{MFd} coincides with the usual evolution considered in mean-field
theory.
It is easy to verify by direct substitution that a noise with the
required correlation function \eq{NoiseCorr} can be obtained as:
\begin{multline}
d\phi_s(\rr,n)=\sum_\kk \left(-\frac{i\,\omega(\kk)}{2{\mathcal N}}\right)^{1/2}\,\left[
dB_\kk\,e^{i\kk\cdot\rr}\,\Big(\sqrt{n}\,\phi(\rr,n-1)-{\bar
  \psi}^*(\rr)\,\phi(\rr,n)\,\Big)+ \right.\\ 
\left.+dB_\kk^*\,e^{-i\kk\cdot\rr}
\,\Big(\sqrt{n+1}\,\phi(\rr,n+1)-{\bar \psi}(\rr)\,\phi(\rr,n)\Big)
\right],
\eqname{MFs}
\end{multline}
with $dB_\kk$ being independent and zero-mean complex random
variables such that:
\begin{eqnarray}
\overline{dB_\kk\,dB_{\kk'}}&=&0 \\
\overline{dB_\kk\,dB^*_{\kk'}}&=&2\,dt\,\delta_{\kk,\kk'}.
\end{eqnarray}
For the lattice here considered, the quasi-momentum $\kk$ is a $D$-component
vector whose $i$th component is equal to $k_i=2\pi n_i /L_i$, $n_i$ being an
integer number $0\leq n_i<{\mathcal N}_i$.
$\omega(\kk)$ is the particle dispersion in the lattice in the absence
of interactions:
\begin{equation}
\hbar \omega(\kk)=J\sum_{i=1}^D \cos(k_i l_i).
\end{equation}
In any practical simulation, a cut-off $N_{\rm max}$ has to be set in
the sums over the site occupation numbers, which physically
corresponds to projecting the dynamics on the 
subspace in which all sites have an occupation number smaller or equal
to $N_{\rm max}$. The error induced by imposing this cut-off is
negligible provided a sufficiently large value of $N_{\rm max}$ is adopted.

\section{Evolution of the statistical error}
\label{sec:StatErr}
As we have discussed in detail in~\cite{Pedag}, formally imposing the correct
value of the averages is not in general sufficient for a practical application:
it is essential to prove the finiteness of the statistical error,
which guarantees the convergence of the Monte Carlo simulation to the
exact result in the limit of an infinite number of independent
realizations. 

As usual, the variance of the statistical error on the state vector of the system is
defined in terms of the stochastically evolving Gutzwiller ansatz
$\ket{G:\phi(t)}$ and the exact state vector
$\ket{\psi(t)}$ as:
\begin{equation}
E(t)=\expect{\big \| \,\ket{G:\phi(t)}-\ket{\psi(t)}\, \big\|^2}.
\eqname{E}
\end{equation}

Using the definition \eq{Average} of the stochastic reformulation, one
can write the evolution of $\ket{G:\phi(t)}$ during an interval $dt$ as: 
\begin{equation}
d\ket{G:\phi(t)}=\frac{dt}{i\hbar}{\mathcal
  H}\ket{G:\phi(t)}+\left.d\ket{G:\phi(t)}\right._{s} 
\end{equation}
where the Ito noise term $\left.d\ket{G:\phi(t)}\right._{s}$ depends
linearly on the stochastic term $d\phi_s(\rr,n)$ of the evolution of the
Gutzwiller wave function. 
Our specific choice of noise terms \eq{MFs} not only leads to the orthogonality
of this Ito term to the Gutzwiller ansatz:
\begin{equation}
\bra{G:\phi}\,\left.d\ket{G:\phi(t)}\right._{s}=0
\end{equation}
but, even more strongly, to a purely orthogonal stochastic 
variation of the state of each site $\rr$:
\begin{equation}
\sum_{n=0}^{\infty} \phi^*(\rr,n)\,d\phi_s(\rr,n)=0.
\eqname{ps}
\end{equation}
Using the fact that the square modulus of the exact state vector $\|
\ket{\psi(t)} \|^2$ is constant and equal to $1$ during the evolution,
the variance of the statistical error can be rewritten as:
\begin{equation}
E(t)=\expect{\Delta(t)}-1
\eqname{E2}
\end{equation}
with:
\begin{equation}
\Delta(t)=\big\|\, \ket{G:\phi(t)} \big\|^2.
\end{equation}
Inserting the explicit form of the stochastic noise term \eq{MFs} and
taking advantage of \eq{ps}, one obtains an upper bound to the
variation of $d\Delta$ in the interval $dt$: 
\begin{equation}
d\Delta=\big\|\left.d\ket{G:\phi}\right._{s}\big\|^2
= \Delta(t) \sum_\rr\frac{\langle d\phi_s(\rr)|d\phi_s(\rr)\rangle}{\langle\phi(\rr)|\phi(\rr)\rangle}
\leq\frac{{\mathcal N} D
J}{\hbar} (N_{\rm max}+1)\,\Delta(t)\,dt
\eqname{dE}
\end{equation}
and therefore on $\Delta(t)$ itself:
\begin{equation}
    \Delta(t)\leq \Delta(0)\,e^{{\mathcal N} D J (N_{\rm max}+1) t/\hbar}.
\end{equation}
This shows that the stochastic trajectories for the Gutzwiller ansatz cannot
escape to infinity in a finite time.
By ensemble averaging, one immediately finds an upper bound to the
variance $E(t)$ of the statistical error \eq{E2}:
\begin{equation}
  \label{eq:error}
  E(t)\leq [E(0)+1]\,e^{{\mathcal N} D J (N_{\rm max}+1) t/\hbar}-1.
\end{equation}
This inequality guarantees that the statistical error of the stochastic
reformulation using Gutzwiller states remains finite at all evolution times.

This result is to be compared with the one we have obtained in~\cite{GPstoch}
for the Fock simple scheme:
\begin{equation}
E(t)\leq [E(0)+1] \, e^{ N U t/ \hbar} -1,
\end{equation}
where $N$ is the number of atoms.
It is apparent how for $N\approx {\mathcal N}$ and $N_{\rm max}\approx
1$, the present scheme is more suited to simulate a system
in the Mott-insulator phase $U\gg J$, for which the interaction energy
dominates over the kinetic energy.

\section{Imaginary time evolution}
\label{sec:ImagTime}
The previous sections have been devoted to the discussion of a stochastic
reformulation of the real-time evolution of the Bose-Hubbard model. 
On the other hand, the density matrix $\rho=\exp(-H/k_B T)$ describing the
thermodynamical equilibrium state of a system at a given temperature
$T$ can be obtained by means of the imaginary-time evolution:
\begin{equation}
\frac{d\rho}{d\tau}=-\frac{1}{2}\big\{H,\rho(\tau)\big\}
\eqname{ImagTime}
\end{equation}
during an imaginary-time interval $\tau=0\rightarrow \beta=1/k_B T$
starting from the identity $\rho(0)={\mathbf 1}$ 
at $\tau=0$. The expectation value of any operator ${\hat O}$ can
be then obtained from $\rho(\beta)$ as:
\begin{equation}
\langle {\hat O} \rangle=\frac{{\textrm Tr}[\rho {\hat O}]}{{\textrm Tr}[\rho]}
\end{equation}
As we have done in~\cite{GPstochT} for the Fock simple scheme, the stochastic
reformulation in terms of a Gutzwiller ansatz originally developed for the
real-time evolution can be easily extended to the imaginary-time evolution. 

It follows from the completeness of coherent states that the identity
matrix can be expanded as:
\begin{equation}
  \label{eq:identity}
  \rho(0)={\mathbf 1}=\frac{1}{\pi^{\mathcal N}}\int\!{\mathcal D}\gamma\,\ket{\textrm{coh:}\gamma}\bra{\textrm{coh:}\gamma}
\end{equation}
in terms of coherent states of the form \eq{Coherent}. The
integration element ${\mathcal D}\gamma$ is defined as usual as:
\begin{equation}
{\mathcal D}\phi=\prod_\rr\,d\,\textrm{Re}\big[\gamma(\rr)\big]\;d\,\textrm{Im}\big[\gamma(\rr)\big].
\end{equation}

According to \eq{identity}, the imaginary-time evolution therefore
starts from initially identical Gutzwiller 
wavefunctions $\phi_{1,2}(\rr,n;0)$ of the coherent state
form \eq{GutzCoh} with a randomly chosen 
$\gamma(\rr)$ and proceeds as described by a pair of stochastic differential equations for
$\phi_{\alpha=1,2}$:
\begin{equation}
d\phi_\alpha(\rr,n)=d\phi_{d,\alpha}(\rr,n)+d\phi_{s,\alpha}(\rr,n).
\end{equation}
Apart from the obvious substitution $\frac{dt}{i\hbar} \rightarrow
-\frac{d\tau}{2}$, the imaginary-time stochastic equations have
exactly the same form as \eq{MFd} and \eq{MFs}. As the Gutzwiller ansatz does not have
in general a well defined number of particles, it is more convenient to use
the Grand-Canonical ensemble with a chemical potential $\mu$, so that
$H={\mathcal H}-\mu {\hat N}$, ${\hat N}$ being the operator giving
the total number of particles. The
deterministic part of the evolution then reads:
\begin{multline}
d\phi_{d,\alpha}(\rr,n)=-\frac{d\tau}{2}\Big[
\Big(\frac{U\,n(n-1)}{2}-\mu\,n\Big)\phi_\alpha(\rr,n)+\frac{J}{2}\sum_{\rr'}\delta_{\rr,\rr'}\al{1}
    \left({\bar
    \psi_\alpha}(\rr')\,\sqrt{n}\,\phi_\alpha(\rr,n-1) +{\bar
    \psi}_\alpha^*(\rr')\,\sqrt{n+1}\,\phi_\alpha(\rr,n+1)\right)\\ 
    -\frac{J}{2{\mathcal N}}\Big(\sum_{\uu,\uu'}\delta_{\uu,\uu'}\al{1} {\bar
    \psi}_\alpha^*(\uu){\bar
    \psi_\alpha}(\uu')\Big)\,\phi_\alpha(\rr,n)\Big]
\end{multline}
and the stochastic part:
\begin{multline}
d\phi_{s,\alpha}(\rr,n)=\sum_\kk
\left(-\frac{\hbar\omega(\kk)}{4{\mathcal N}}\right)^{1/2}\,\left[
dB_\kk\,e^{i\kk\cdot\rr}\,\Big(\sqrt{n}\,\phi(\rr,n-1)-{\bar
  \psi}_\alpha^*(\rr)\,\phi_\alpha(\rr,n)\,\Big)+ \right.\\ 
\left.+dB_\kk^*\,e^{-i\kk\cdot\rr}
\,\Big(\sqrt{n+1}\,\phi_\alpha(\rr,n+1)-{\bar
  \psi}_\alpha(\rr)\,\phi_\alpha(\rr,n)\Big) 
\right],
\end{multline}
with $dB_\kk$ being independent and zero-mean complex random
variables such that:
\begin{eqnarray}
\overline{dB_\kk\,dB_{\kk'}}&=&0 \\
\overline{dB_\kk\,dB^*_{\kk'}}&=&2\,d\tau\,\delta_{\kk,\kk'}.
\end{eqnarray}
In the practical simulation presented below,
the $dB_{\kk}$ have a Gaussian distribution.
For the left and right Gutzwiller ansatz $\alpha=1,2$,
${\bar \psi}_\alpha(\rr)$ is defined as the mean-field corresponding to the wavefunction
$\phi_\alpha(\rr,n)$:
\begin{equation}
{\bar \psi}_\alpha(\rr)=\frac{\braopket{G:\phi_\alpha}{\Psih(\rr)}{G:\phi_\alpha}}
{\braket{G:\phi_\alpha}{G:\phi_\alpha}}.
\end{equation}
The finiteness of the statistical error at any time of the imaginary-time
evolution can be proven by means of the same arguments invoked above for the
case of a real-time evolution. The evolution of the state vector
during $d\tau$ is given by:
\begin{equation}
d\ket{G:\phi(\tau)}=-\frac{d\tau}{2}H\ket{G:\phi(t)}+\left.d\ket{G:\phi(\tau)}\right._{s}.
\end{equation}
As the spectrum of the Bose-Hubbard Hamiltonian $H$ is
bounded from below in the presence of the cut-off at $N_{\rm max}$, an upper
bound can be set to the growth rate of
$\|\,\ket{G:\phi(\tau)}\,\|^2$, which guarantees that the 
statistical error remains finite at any time.
\begin{figure}[htbp]
\begin{center}
\psfig{figure=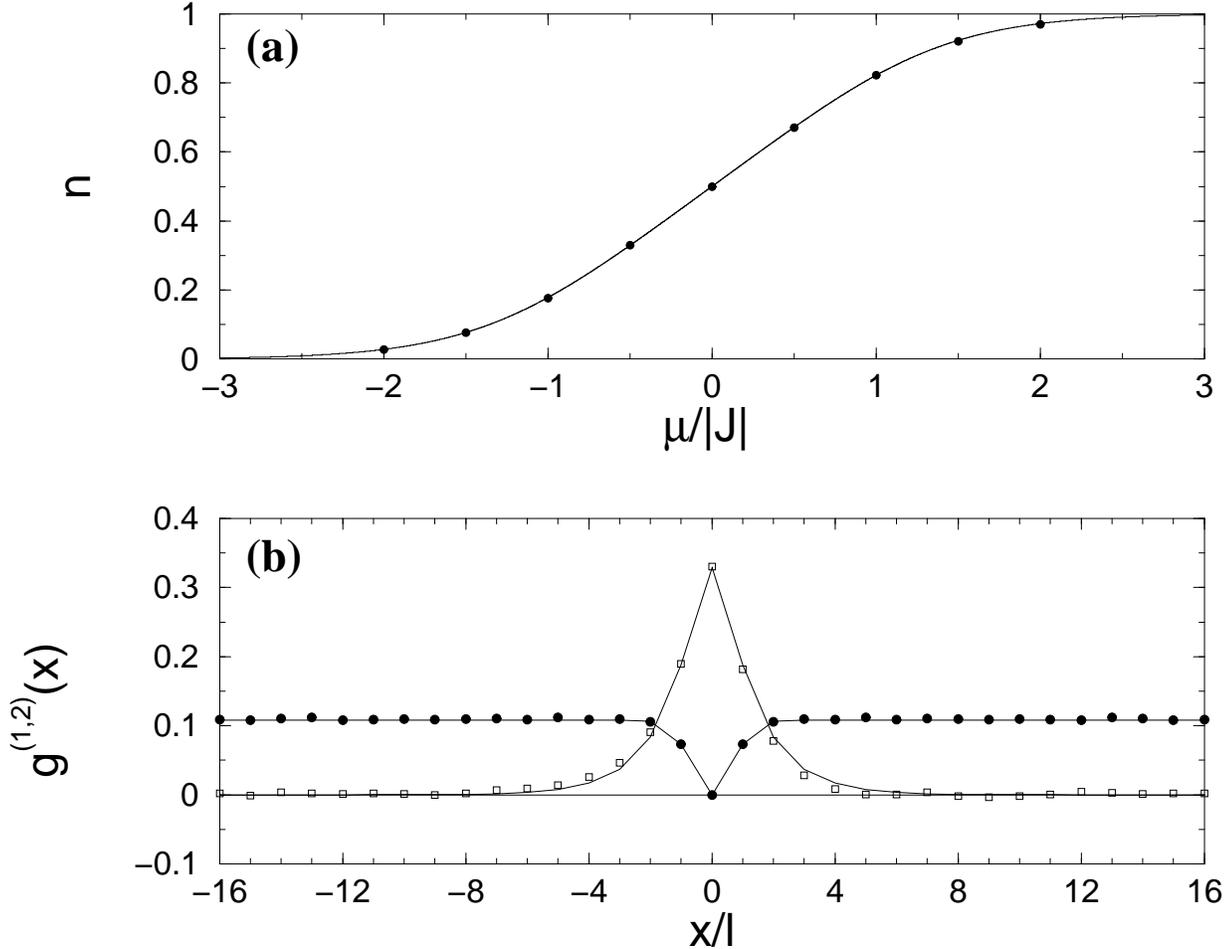,width=16cm}
\end{center}
\caption{For a 1D lattice with 32 sites and the impenetrable boson regime
$U=+\infty$, (a) mean filling factor $n$ 
as a function of the chemical potential, (b) for $\mu=-|J|/2$, 
first order correlation function
$g^{(1)}(x)=\langle \Psihd(x)\Psih(0)\rangle$ (open squares) 
and second order correlation
function $g^{(2)}(x)=\langle\Psihd(x)\Psihd(0)\Psih(0)\Psih(x)\rangle$
(black disks) as function of the coordinate $x$. 
In (a) and (b), the solid lines are the analytical results obtained by
means of fermionization tecniques as detailed in the Appendix.
In both panels, the temperature is related to the (here negative) hopping constant
by $|J|/k_B T = 9/4$. In (b) the difference between the chemical potential
and the single particle ground state energy $J$ is $(9/8) k_B T$ so that
the fermionic analog of the Bose system is on the border of being degenerate.
The position $x$ is in units of the lattice spacing $l$.}
\end{figure}

In the figure, we present a first, simple application of the method, 
the observation of the finite temperature
cross-over from the discrete Tonks gas to the Mott phase in a system
of impenetrable spinless bosons. This system for which $U=+\infty$ was totally out of reach for our 
previous algorithm with Hartree states \cite{GPstochT} but can be
easily treated by the stochastic Gutzwiller algorithm by simply
setting $N_{\rm max}=1$.
We introduce the name `discrete Tonks gas' to refer to a phase of impenetrable bosons
on a lattice with a filling factor much smaller than unity; as the mean distance between bosons
is then much larger than the lattice spacing, we expect this phase to have large scale properties 
very close to the ones of the continuous model, hence the name.
The Mott phase corresponds to the opposite limit of a unit filling factor.

\section{Several spin states}
\label{sec:Spin}
All the discussion of the previous sections can be easily extended to the case
in which the particles can be in $s$ different spin states.
The Bose-Hubbard Hamiltonian \eq{Hamilt} is immediately generalized as: 
\begin{equation}
{\mathcal H}={\mathcal H}_U+{\mathcal
  H}_J=\sum_\rr\sum_{\sigma,\sigma'=1}^s\frac{U_{\sigma\sigma'}}{2}
  \Psihd(\rr,\sigma)\Psihd(\rr,\sigma')\Psih(\rr,\sigma')\Psih(\rr,\sigma)+\sum_{\rr,\rr'}\sum_{\sigma=1}^s
  \frac{J_\sigma}{2}\,\delta_{\rr,\rr'}\al{1}
\Psihd(\rr,\sigma)\Psih(\rr',\sigma), 
\eqname{HamiltSigma}
\end{equation}
$\Psih(\rr,\sigma)$ being the destruction operators for a particle at the site
$\rr$ in the spin state $\sigma$.
The matrix $U_{\sigma,\sigma'}$ quantifies the on-site interactions between
atoms respectively in the $\sigma$ and $\sigma'$ states. The hopping is
assumed not to affect the spin degrees of freedom and is quantified by the
parameters $J_\sigma$.

The straightforward generalization of the Gutzwiller ansatz \eq{Gutzwiller} to
the case of several spin states has the form:
\begin{equation}
\ket{G:\phi}=\prod_\rr{\sum_{n_1\ldots n_s=0}^{\infty}
  \phi(\rr,\nn) \prod_{\sigma=1}^s\frac{\Psihd(\rr,\sigma)^{n_\sigma}}{\sqrt{n_\sigma !}}}\,\ket{0},
\label{eq:GutzwillerSigma}
\end{equation}
where the Gutzwiller wavefunction $\phi(\rr,\nn)$ now depends on the spatial
coordinate $\rr$ and on the $s$-components vector $\nn=(n_1,\ldots,n_s)$ formed
by the occupation numbers of each spin state.

As in the spinless case, both the on-site interactions and the mean-field
contribution to the hopping term are taken into account by the deterministic
evolution of $\phi(\rr,\nn)$. The fluctuations around mean-field are described
by the stochastic noise term.

For the case of a real-time evolution, the deterministic term reads:
\begin{multline}
d\phi_d(\rr,\nn)=
\frac{dt}{i\hbar}\Big[\sum_{\sigma,\sigma'}\frac{U_{\sigma\sigma'}\,n_\sigma
 (n_{\sigma'}-\delta_{\sigma,\sigma'}) }{2}\,\phi(\rr,\nn)+\\ +\sum_{\sigma,\rr'} \frac{J_\sigma}{2}
\delta_{\rr,\rr'}\al{1}
    \left({\bar
    \psi}(\rr',\sigma)\,\sqrt{n_\sigma}\,\phi(\rr,\nn-\ee_\sigma)+{\bar
    \psi}^*(\rr',\sigma)\,\sqrt{n_\sigma+1}\,\phi(\rr,\nn+\ee_\sigma)\right)-\Big(\sum_{\uu,\uu',\sigma}\frac{J_\sigma}{2{\mathcal N}}\delta_{\uu,\uu'}\al{1} 
    {\bar 
    \psi}^*(\uu,\sigma){\bar \psi}(\uu',\sigma)\Big)\,\phi(\rr,\nn)\Big].
\eqname{MFdSpin}
\end{multline}
All the sums over spin states run over $\sigma=1\ldots s$ and
$\ee_\sigma$ is defined as the $s$-component unit vector whose
$\sigma$-th component is $1$ and all the others 
are vanishing. The mean-field ${\bar \psi}(\rr,\sigma)$ is defined as:
\begin{equation}
{\bar \psi}(\rr,\sigma)=
\frac{\braopket{G:\phi}{\Psih(\rr,\sigma)}{G:\phi}}
{\braket{G:\phi}{G:\phi}}.
\end{equation}
The noise term can be written as:
\begin{multline}
d\phi_s(\rr,n)=\sum_{\kk,\sigma}
\left(-\frac{i\,\omega(\kk,\sigma)}{2{\mathcal N}}\right)^{1/2}\,\left[
dB_{\kk,\sigma}\,e^{i\kk\rr}\,\Big(\sqrt{n_\sigma}\,\phi(\rr,\nn-\ee_\sigma)-
{\bar
  \psi}^*(\rr,\sigma)\,\phi(\rr,\nn)\,\Big)+ \right.\\ 
\left.+dB_{\kk,\sigma}^*\,e^{-i\kk\rr}
\,\Big(\sqrt{n_\sigma+1}\,\phi(\rr,\nn+\ee_\sigma)-{\bar \psi}(\rr,\sigma)\,
\phi(\rr,\nn)\Big)
\right],
\eqname{MFsSpin}
\end{multline}
in terms of the independent and zero-mean complex random
variables $dB_{\kk,\sigma}$ such that:
\begin{eqnarray}
\overline{dB_{\kk,\sigma}\,dB_{\kk',\sigma'}}&=&0 \\
\overline{dB_{\kk,\sigma}\,dB^*_{\kk',\sigma'}}&=&2\,d\tau\,\delta_{\kk,\kk'}\,\delta_{\sigma,\sigma'}.
\end{eqnarray} 
$\omega(\kk,\sigma)$ describes the particle dispersion in the lattice
in the absence of interactions:
\begin{equation}
\hbar \omega(\kk,\sigma)=J_\sigma\sum_{i=1}^d \cos(k_i l_i).
\end{equation}
As in the previous section, these equations can be extended to the
imaginary-time evolution simply by replacing  $\frac{dt}{i\hbar} \rightarrow
-\frac{d\tau}{2}$. The finiteness of the statistical error can be
proven in the same way as previously done for the spinless case.

\section{Conclusions}
\label{sec:Conclu}
In this paper, we have introduced a reformulation of the
bosonic many-body problem on a lattice in terms of an stochastic
Gutzwiller ansatz and we have proved that the statistical dispersion
remains finite at all times in the presence of a cut-off on the number
of particles per lattice site. This reformulation allows to perform efficient Monte Carlo
simulations for strongly correlated systems like the Mott phase or the Tonks gas.
This new method can also be used to study hard sphere bosons in arbitrary
spatial dimensions, a situation which was totally out of reach for our
previous stochastic Hartree ansatz.

\begin{acknowledgments}
We acknowledge useful discussions with M. Fleischhauer, L. Plimak, P. Drummond,
K. Kheruntsyan, P. Zoller, A. Recati, D. Gangardt and T. Jolic\oe ur.
I.C. acknowledges a Marie Curie grant from the EU under contract number
HPMF-CT-2000-00901.  We are grateful to r\'egion Ile de France for financial support.
Laboratoire Kastler Brossel is a Unit\'e de
Recherche de l'\'Ecole Normale Sup\'erieure et de l'Universit\'e Paris
6, associ\'ee au CNRS.

\end{acknowledgments}

\appendix*
\section{Analytical calculation of the correlation functions of
impenetrable bosons}

The correlation functions of a gas of impenetrable bosons in one
dimension can be computed by mapping the bosonic system onto a
fermionic system. The bosonic operators $\Psih(x)$ on the lattice are written
in terms of fermionic operators $\Psih_F(x)$ by means of the
Jordan-Wigner transformation~\cite{Cazalilla,Larkin}: 
\begin{equation}
\Psih(x)=\exp\Big(i\pi\sum_{x'=0}^{x-l}\Psihd_F(x')\Psih_F(x')\Big)\Psih_F(x).
\eqname{J-W}
\end{equation}
In this way, the Hamiltonian \eq{Hamilt} for the impenetrable ($U=+\infty$)
Bose gas reduces to a free-fermion Hamiltonian of the form:
\begin{equation}
{\mathcal
H}=\frac{J}{2}\Big[\sum_{x=0}^{L-2l}\Psihd_F(x+l)\Psih_F(x)+(-1)^{{\hat
N}-1}\Psihd_F(0)\Psih_F(L-l)+\textrm{h.c.}\Big].
\eqname{Hamilt_F}
\end{equation}
Given the specific ordering given to sites in \eq{J-W}, it is in fact
necessary to isolate in \eq{Hamilt_F} the term corresponding to 
the tunneling between the last and the first site of the
lattice, whose amplitude changes sign depending on the parity of the
total number of particles present in the system.
This peculiarity of the periodic boundary conditions is
well-known in the frame of exactly solvable one-dimensional models,
see e.g. \cite{Lieb}.

For any value of the total number of particles, the Hamiltonian
\eq{Hamilt_F} is quadratic in the Fermi field operators and describes
a homogeneous system of non-interacting fermions. The single particle
states are plane waves of wavevector $k$ and their energy is given by:
\begin{equation}
\hbar\omega(k)=J\cos(k l),
\eqname{omega_k}
\end{equation}
but the quantization condition on $k$ depends on the parity of $N$.
For odd values of $N$, the Hamiltonian \eq{Hamilt_F} reduces to the
usual one for a one-dimensional lattice with tunneling constant $J$,
so that periodic boundary conditions on the single-particle
wavefunctions apply and the allowed values for $k$ are given by:
\begin{equation}
k=\frac{2 \pi}{L} n
\eqname{k_odd}
\end{equation}
in terms of the integer $0\leq n < {\mathcal N}$. On the other hand, for even
values of $N$, anti-periodic boundary conditions are to be applied, so
that $k$ is now given by:
\begin{equation}
k=\frac{2 \pi}{L}(n+\frac{1}{2}).
\eqname{k_even}
\end{equation}

As the total number of particles $N$ commutes with the Hamiltonian, we
can write the Grand-Canonical density operator of our system as the sum of the
contributions of respectively the odd and the even values of $N$:
\begin{equation}
\rho=\rho_{\rm odd}+\rho_{\rm even},
\eqname{rho}
\end{equation}
where:
\begin{equation}
\rho_{\rm even}=\frac{1}{2}\Big[ e^{-\beta H_{\rm even}}+e^{-\beta
H_{\rm even}+i\pi {\hat N}}\Big]
\eqname{rho_even}
\end{equation}
and 
\begin{equation}
\rho_{\rm odd}=\frac{1}{2}\Big[ e^{-\beta H_{\rm odd}}-e^{-\beta
H_{\rm odd}+i\pi {\hat N}}\Big],
\eqname{rho_odd}
\end{equation}
which involve modified Grand-Canonical density operators with $\beta \mu$
replaced by $\beta \mu+i\pi$.

The odd-$N$ Hamiltonian $H_{\rm odd}$ reads:
\begin{equation}
H_{\rm
odd}=\frac{J}{2}\Big[\sum_{x=0}^{L-2l}\Psihd_F(x+l)\Psih_F(x)+\Psihd_F(0)\Psih_F(L-l)+\textrm{h.c.}
\Big]-\mu {\hat N} 
\eqname{H_odd},
\end{equation}
while the even-$N$ one reads:
\begin{equation}
H_{\rm even}=\frac{J}{2}\Big[\sum_{x=0}^{L-2l}\Psihd_F(x+l)\Psih_F(x)-\Psihd_F(0)\Psih_F(L-l)+\textrm{h.c.}\Big]-\mu {\hat N}
\eqname{H_even}.
\end{equation}

In order to obtain the expectation value $\langle {\hat O}\rangle$ of
an arbitrary observable ${\hat O}$, we have to calculate quantities such as
$\textrm{Tr}[\sigma]$ and $\langle {\hat O}
\rangle_\sigma=\textrm{Tr}[{\hat O}
\sigma]/\textrm{Tr}[\sigma]$ where $\sigma$ is one of the four
density operators appearing on the right-hand side of \eq{rho_even} and
\eq{rho_odd}.
As the $\sigma$ are Gaussian operators of the form $\exp(A)$ with $A$
being a quadratic (but non-necessarily hermitian) operator in the
Fermi field operators $\Psih_F(x)$, one has:
\begin{equation}
\textrm{Tr}[\sigma]=\prod_k \big(1+\exp(a_k)\big)
\eqname{trace_sigma}
\end{equation}
with the $a_k$'s being the single particle eigenvalues of
$A$. Depending whether one considers the first or the second term in
\eq{rho_even}, \eq{rho_odd},
the eigenvalues $a_k$ are equal to $-\beta(\hbar \omega(k)-\mu)$,
or to $-\beta(\hbar
\omega(k)-\mu)+i\pi$, with the values of $k$ corresponding to the
relevant quantization conditions \eq{k_odd} or \eq{k_even}.

The calculation of $\langle {\hat O} \rangle_\sigma=\textrm{Tr}[{\hat O}
\sigma]/\textrm{Tr}[\sigma]$ has to be performed using the
Jordan-Wigner mapping \eq{J-W} to write the observable in terms of the
fermionic operators and then applying Wick's theorem, whose validity
can be proven to extend to the present case of a non-hermitian
quadratic Hamiltonian.
In particular, the first order correlation function $g\al{1}(x)$ in
one of the density operators $\sigma$ can be shown~\cite{Larkin,YvanHouches} to be given by:
\begin{equation}
g\al{1}(x\neq0)=\big\langle\Psihd(0)\Psih(x)\big\rangle_\sigma=\left.\frac{1}{2}\det\Big[2\big\langle \Psihd_F(y)\Psih_F(y'+l)
\big\rangle_\sigma-\delta_{y,y'+l}\Big]\right|_{y,y'=0\ldots x-l}.
\end{equation}
Note that we have added a global factor $1/2$ which was missing in \cite{Larkin}.

\end{document}